\documentclass[traditabstract]{aa}
\usepackage{graphicx}                    
\usepackage{bm}
\usepackage{color}                       
\usepackage{url}                         

\renewcommand{\vec}[1]{\mbox{\boldmath$#1$}}
\def\gsim{\lower.4ex\hbox{$\;\buildrel >\over{\scriptstyle\sim}\;$}} 
\def\lsim{\lower.4ex\hbox{$\;\buildrel <\over{\scriptstyle\sim}\;$}}

\def \Om  {{\it \Omega}}
\begin{document}



\title{The angular momentum transport by  unstable  toroidal magnetic 
fields}
\titlerunning{Viscosity by unstable magnetic toroidal fields}
%
\author{G.~R\"udiger \and  M. Gellert \and F. Spada  \and  I. Tereshin}


%
  \institute{Leibniz-Institut f\"ur Astrophysik Potsdam, An der Sternwarte 16, D-14482 Potsdam, Germany,
                     email:  gruediger@aip.de
		     }

\date{Received; accepted}
 
\abstract{We demonstrate with a nonlinear MHD code that {angular momentum can be transported} due to the magnetic instability of toroidal fields under the influence of differential rotation, and that the resulting effective viscosity may be high enough to explain the almost rigid-body rotation observed in radiative stellar cores. Only stationary current-free fields and only those combinations of rotation rates and magnetic field amplitudes  { which provide maximal numerical values of the viscosity are considered}. We find that the dimensionless { ratio of the effective over molecular viscosity}, $\nu_{\rm T}/\nu$,  linearly grows with the  Reynolds number of the rotating fluid multiplied with the square-root of the magnetic Prandtl number -- which is of order unity for the considered red sub-giant KIC 7341231. 

For the considered interval of magnetic Reynolds numbers -- which is restricted by numerical constraints of the nonlinear MHD code -- there is a remarkable influence of the magnetic Prandtl number on the { relative importance of the} contributions of the Reynolds stress and the Maxwell stress to the total viscosity, which is magnetically dominated { only} for $\rm Pm\gsim 0.5$. We also find { that} the magnetized { plasma behaves as a non-Newtonian fluid}, i.e. the resulting effective viscosity depends on the shear in the rotation law. The decay time of { the differential rotation} thus depends on its shear and becomes { longer and longer} during the spin-down of a stellar core.}


%

\maketitle
%
\section{Introduction}
 Model calculations for the formation of red giants without turbulent or magnetic angular momentum transport lead to rather steep radial profiles of the angular velocity in the innermost core of the star. Ceillier et al. (2012) report a theoretical (quasi-Keplerian) profile $\Om \propto R^{-q}$ with $q\simeq 1.6$ for the low-mass red giant KIC 7341231. Note that the exponent $q=1$ would describe  a quasi-galactic rotation profile with $U_\phi=$ const while $q=2$ represents the rotation law for uniform angular momentum $\Om R^2\approx$ const. The {\em Kepler} data, however, lead to much flatter rotation laws of the observed red giants: the cores of several sub-giants and young red giants  seem to rotate only (say) five times faster than the outer convection zone (Deheuvels et al. 2012, Deheuvels et al. 2014).  
 Eggenberger et al. (2012)  argue that only an additional viscosity of $3 \times 10^4$ cm$^2$/s may explain  the unexpectedly flat internal rotation law of the more massive red giant KIC 8366239. The outward flux of angular momentum { due to this enhanced} viscosity,{ which exceeds} the molecular value by a factor of $\nu_{T}/\nu \approx 500$, suffices to produce the observed spin-down of the inner radiative core. R\"udiger \& Kitchatinov (1996) needed just this viscosity value to produce by Maxwell stress the high degree of uniformity of the internal solar rotation, derived from helioseismologic measurements down to $0.15 R_\odot$.

Rotation laws with $q<2$ are hydrodynamically stable. Under the presence of toroidal fields, however, they become unstable against nonaxisymmetric disturbances if the amplitude of the toroidal field is high enough but does not exceed  $\Om_{\rm A}\simeq \Om$ with  $\Om_{\rm A}=B_\phi/\sqrt{\mu_0 \rho R^2}$ as the Alfv\'{e}n frequency for incompressible fluids. This nonaxisymmetric instability even exists for toroidal fields which are current-free in the considered domain. Because of this force-free character of the magnetic field, the instability has been called the `azimuthal magnetorotational instability' (AMRI). Within a cylindrical setup, AMRI has been studied in the linear approximation, but also by nonlinear simulations (R\"udiger et al. 2014). The consequences of both compressibility and heat transport (see Spruit 2002) cannot be studied with the present model. We know, however, that these influences become negligible for strong fields. One also can show that, with thermodynamics included, the radial components of flow and field are strongly damped, so that the resulting angular momentum transport should be reduced by the `negative buoyancy'. The viscosity values derived in the present paper are thus maximum values. If they are not high enough for $\nu_{\rm T}/\nu\simeq 500$ then the concept of the instability of magnetic fields in the stellar interior is proven as not working.

An important  basis for realistic numerical simulations is the knowledge of the magnetic Prandtl number
 \begin{equation}
 {\rm Pm}=\frac{\nu}{\eta}, 
 \label{Pm}
\end{equation}
 where $\nu$ is the molecular viscosity of the fluid and $\eta$ its magnetic diffusivity. So far, numerical nonlinear simulations are only possible for $\rm Pm$ exceeding (say) 0.01. The magnetic Prandtl number of the plasma inside main-sequence stars, however, is smaller (see Brandenburg \& Subramanian 2005). We have thus first to probe the value of $\rm Pm$ in the radiative interiors of the considered red {\em Kepler} stars.

 \section{The stellar model of KIC 7341231}
 \label{sect1} 
A model for the star KIC 7341231 is calculated using the Yale Rotational stellar Evolution Code (YREC) in its nonrotational configuration (Demarque et al. 2008). 
The code uses up-to-date input physics, such as OPAL 2005 equations of state (Rogers $\&$ Nayfonov 2002) and OPAL opacities (Iglesias $\&$ Rogers 1996). 
The treatment of the atmospheric boundary conditions is based on the Eddington grey $T$-$\tau$ relation; convection is described according to the mixing length theory (B\"ohm-Vitense 1958), with the mixing length parameter set to $\alpha_{\rm MLT} = 1.82$, our solar-calibrated value. 
Our choice of the stellar parameters (e.g. $M=0.84 \, M_{\odot}$, $Z/X = -1.0$, age $\approx$ 12 Gyr, etc.) is based on the best-fitting model of Deheuvels et al. (2012, see their table 3). 
Our model of KIC 7341231 has effective temperature $T_{\rm eff} = 5380$ K, radius $R = 2.66 R_\odot$, and a fractional radius at the bottom of the convective envelope of $r_{\rm bce} = 0.34$ at the age of 11.8 Gyr. 
\begin{figure}[t!]
\begin{center}
\includegraphics[width=\columnwidth]{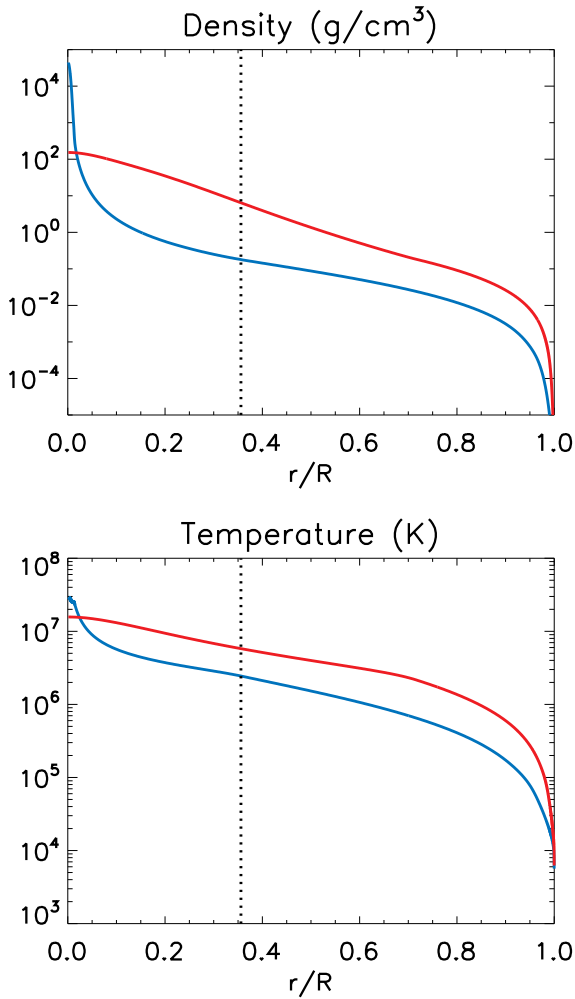}
\end{center}
\caption{The stellar model of KIC 7341231 (blue) { compared} to a standard solar model (red). The vertical dashed line gives the outer boundary of the radiative stellar core { of the red sub-giant}.
 }
\label{model}
\end{figure}

\begin{figure}[t!]
\begin{center}
\includegraphics[width=\columnwidth]{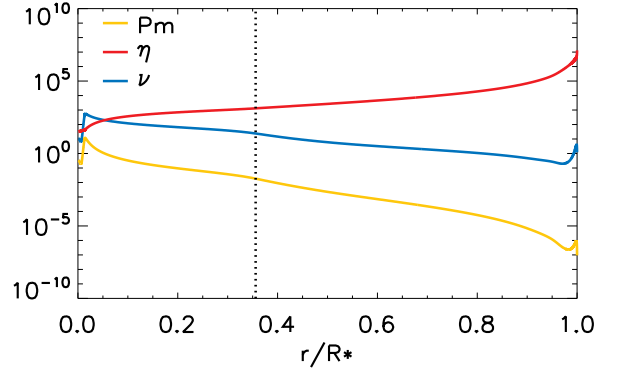}
\end{center}
\caption{The diffusion coefficients from the stellar model of KIC 7341231. The vertical dashed line gives the outer boundary of the radiative stellar core. Note that the geometric average of the two diffusivities ${\bar \eta} =\sqrt{\nu\eta}\approx 100$\ cm$^2$/s.
 }
\label{prandtl}
\end{figure}

The run of the viscosity $\nu$ and magnetic diffusivity $\eta$ in the interior of this model, given by
\begin{eqnarray}
\nu &=& \nu_{\rm mol} + \nu_{\rm rad} = 1.2 \cdot 10^{-16} \frac{T^{5/2}}{\rho} + 2.5 \cdot 10^{-25} \frac{T^4}{\kappa \rho},
\nonumber\\
\eta &=& 10^{13} T^{-3/2}
\end{eqnarray}
(all in c.g.s. units), is shown in Fig. \ref{prandtl}.   
The basic result is that a characteristic value for the microscopic  magnetic diffusivity  is 10$^3$ cm$^2$/s, while the magnetic Prandtl number {\em varies between $0.1$ and $10$}. The microscopic magnetic diffusivity is thus of the order of that of the Sun, while the microscopic viscosity is much higher. It makes thus sense to focus the attention to the results as a function of the magnetic Reynolds number rather than the ordinary Reynolds number. 
For { different} stellar models, therefore, the magnetic Reynolds number mainly { encompasses} the differences of the angular velocity of the inner rotation. 
 The magnetic Reynolds number of the solar core is about $5\cdot 10^{12}$ and its magnetic Prandtl number is $5\cdot 10^{-3}$.

\section{Numerical setup and the calculation of the effective viscosity}
After the results of the above stellar structure calculations it is thus reasonable to perform simulations with magnetic Prandtl numbers between 0.1 and unity. Another approximation should also be mentioned, as the  calculations focus to the  angular momentum  in the equatorial region. After the Taylor-Proudman theorem for fast rotation the $\Om$ forms cylindrical isolines, $\Om=\Om(R)$. More simplifying, the axial structure of the toroidal magnetic field belts is neglected so that also $B_\phi=B_\phi(R)$. The resulting viscosity values certainly overestimate the more realistic corresponding values  for a spherical model.

We consider the same cylindrical setup used by R\"udiger et al. (2014), to which the reader is referred for further details. 
Cylindrical coordinates $(R,z,\phi)$ are adopted, with the $z$ axis coincident with the axis of the bounding cylinders, of radii $R_{\rm in}$ and $R_{\rm out}$, respectively. The cylinders are assumed as rotating with different frequencies so that differential rotation can easily be modeled.

 An important detail of the calculations is the  scaling of the instability for small magnetic Prandtl numbers. For rotation laws with $q\simeq 2$, the bifurcation map scales with the Reynolds number ${\rm Re}$ and the Hartmann number ${\rm Ha}$,
 \begin{equation}
{\rm Re}= \frac{\Om_{\rm in} \, R_0^2}{\nu},    \ \ \ \ \ \ \ \ \ \  \ \ \ \ \  {\rm Ha} = \frac{B_{\rm in} \, R_0}{\sqrt{\mu_0 \rho\nu\eta}} ,
\label{Re}
\end{equation}  
 while for more flat rotation laws  the corresponding parameters are the magnetic Reynolds number ${\rm Rm=Pm \, Re}$ and Lundquist number 
 ${\rm S}=\sqrt{{\rm Pm}} {\rm Ha}$.
The radial scale is $R_0=\sqrt{R_{\rm in}(R_{\rm out}-R_{\rm in})}$.

\begin{figure}[t!]
\begin{center}
\includegraphics[width=9cm]{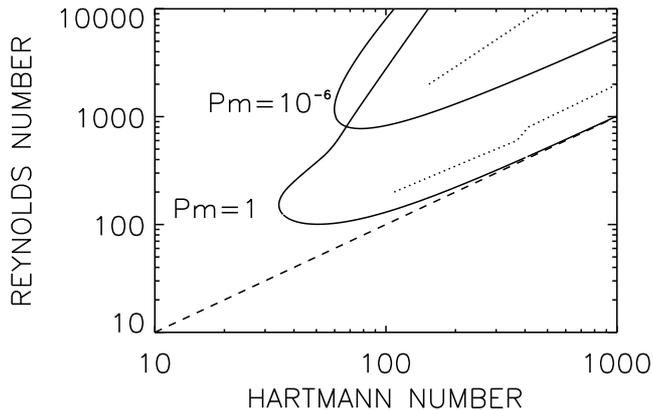}
\end{center}
\caption{The curves of marginal instability for $\Om\propto 1/R^2$ if $r_{\rm in}=0.5$, i.e. $\mu=0.25$. For small magnetic Prandtl numbers, the curves scale with $\rm Re$ and $\rm Ha$. For $\rm Pm=1$, the critical Reynolds number is (only) 100. The dotted lines mark the line of maximum growth rates. The dashed line represents for $\rm Pm=1$ the lower limit of the condition (\ref{inst}).
 }
\label{fig1}
\end{figure}

We imagine the formation of the toroidal field as due to the induction by the differential rotation from a fossil poloidal field $B_{\rm p}$; hence $B_\phi \lsim {\rm Rm} \, B_{\rm p}$, or what is the same, $\Om_{\rm A}/\Om \lsim S_{\rm pol}$ with $S_{\rm pol}$ the Lundquist number { of the poloidal field}:
\begin{equation}
{\rm S}_{\rm p}= \frac{B_{\rm p} \, R_0}{\sqrt{\mu_0\rho} \eta} .
\label{Lund}
\end{equation}
{Thus}, with ${\rm S}_{\rm p}$ of order unity, one finds that always
\begin{equation}
\Om\gsim \Om_{\rm A}.
\label{inst}
\end{equation} 
Note also that  ${\rm S}_{\rm p}\simeq 1$ is already fulfilled with rather weak poloidal fields, { of the order of} $\mu$G.
For the rotation law at the Rayleigh limit ($\mu=0.25$)  and the current-free magnetic background field $B_\phi\propto 1/R$  the flow ${U_\phi}$ and the field $ {B_\phi}$ have the same dependence on the radius. Chandrasekhar (1956) has demonstrated  that all {\em ideal} MHD systems with  $\vec{U}=\vec{B}$ are stable. In contrast, Fig. \ref{fig1} demonstrates that  for nonideal fluids an instability exists for $\Om_{\rm A}= \Om$. Moreover, if the Reynolds number of the fluid exceeds a critical value of (only) 100 then an extended area of instability exists above this line fulfilling just the condition (\ref{inst}).

Tayler (1957, 1973) showed that toroidal fields with large enough amplitude become unstable against nonaxisymmetric disturbances, { in the absence of rotation}. There is a local criterion  for instability written in cylinder coordinates, i.e.
\begin{equation}
\frac{{\rm d}}{{\rm d}R} (R \, B_\phi^2) > 0,
\label{1}
\end{equation}
which is  fulfilled by the magnetic field $B_\phi \propto R$, due to a homogeneous axial electric current. If an electric current flows unbounded in the axial direction, the critical Hartmann number
$
{\rm Ha}_{\rm out} = {B_{\rm out} \, R_0}/{\sqrt{\mu_0 \rho\nu\eta}}
$
is about $20$ which has indeed been realized { in an experiment (Seilmayer et al. 2012, R\"udiger et al. 2012)}. Numerical simulations of this  magnetic {configuration} showed, however, that the eddy viscosity resulting from this instability remains  small compared with its microscopic value.

If the toroidal field is of the vacuum type, it runs as $B_\phi \sim 1/R$. Here the symmetry axis, $R\to 0$, where the electric current flows, must be excluded from the domain of calculations. This field, alone, is stable { according to} the criterion (\ref{1}). The field, however, destabilizes a differential rotation which itself is stable in the hydrodynamical regime.The  energy source of this instability is the differential rotation, which via AMRI drives a nonaxisymmetric magnetic instability pattern; the magnetic energy dominates the energy in the kinetic fluctuations. The basic saturation mechanism of the instability is the turbulence-induced decay of the rotation law. To this end, angular momentum must be transported into the direction of slow rotation, i.e.
\begin{equation}
\langle u_R u_\phi\rangle-\frac{1}{\mu_0\rho}\langle b_R b_\phi\rangle = -\nu_{\rm T} \, R \frac{{\rm d}\Om}{{\rm d}R}.
\label{uRuphi}
\end{equation}
Here  $ \vec{u}$ and $ \vec{b}$ are  the fluctuations of the velocity and magnetic field with their  average values $\vec{U}$ and $ \vec{B}$.

The calculation of the effective viscosity $\nu_{\rm T}$ is thus possible by the calculation of the cross-correlations $\langle u_R u_\phi\rangle$ and  $\langle b_R b_\phi\rangle$. Note that for a magnetic-dominated MHD turbulence the  relation 
\begin{equation}
\nu_{\rm T}\simeq \frac{1}{3} \frac{\langle \vec{b}^2\rangle}{\mu_0 \rho} \tau_{\rm corr}
\label{nuT}
\end{equation}
holds (Vainshtein \& Kichatinov  1983), which may be used as an estimation by replacing the unknown correlation time $\tau_{\rm corr}$ with the growth time  $1/\omega_{\rm gr}$, {where} $\omega_{\rm gr}$ is the maximal  growth rate of the instability (see Spruit 2002).  The maximum growth rates lead to the shortest time-scale in the system. One can indeed show that for the magnetic instability a linear relation of the turnover frequency $\omega_{\rm turn}=k \sqrt{\langle u_z^2}$  and the growth rate exists (Fig. \ref{turnover}). In the sense of the mixing length arguments we have only to put equal the turnover time and the correlation time. This is insofar a striking result as it allows to express the saturated hydrodynamic energy by (the square of) the linear phase velocity. 

R\"udiger et al. (2014) have shown that for ${\rm Pm}=1$ the unstable $m=1$ modes grow with
$
{\omega_{\rm gr}}/{\Om_{\rm in}}\propto {\rm Re}.
$
One also finds for the magnetic energy
$
{\langle \vec{b}^2\rangle}/{B_{\rm in}^2}\propto  {\rm Re}.
$
 Hence,
\begin{equation}
\frac{\nu_{\rm T}}{\nu}\propto  \frac{{\rm Ha}^2}{{\rm Re}}
\label{nuthalbe}
\end{equation}
which for AMRI where  the maximal growth rates  are located at a line  with  ${\rm Ha}\propto {\rm Re}$ (see Fig. \ref{fig1}) leads to a linear relation  ${\nu_{\rm T}}/{\nu}\propto  {{\rm Re}}$. The question is how this relation saturates by the magnetic feedback and whether this maximal effective viscosity is large enough to explain the above mentioned {\em Kepler} observations.
The results of our { numerical experiments indeed confirm that} $\nu_{\rm T} \propto \Om_{\rm in} \, R_0^2$. The coefficient in this relation must be fixed with numerical calculations. With our numerical tools, however, we could not find the saturation effect so that the question after the maximally possible viscosities requires better numerical qualities. What we have nevertheless shown is that indeed i) the Maxwell stress by the fluctuations of the azimuthal magnetorotational instability mainly forms the effective viscosity and ii) the viscosity for slow rotation  runs linearly 
with magnetic Reynolds number.

We shall simulate  two different rotation laws, i.e. the two limits $\Omega \propto 1/R^2$ and $\Om \propto 1/R$, { bracketing} the rotation law { obtained in modeling sub-giant stars} without enhanced transport of angular momentum, i.e. $\Om \propto R^{-1.6}$ { (Ceillier et al. 2012)}.

We start with the steep rotation law $\Omega\propto 1/R^2$, which is one of the two basic solutions for the stationary rotation law between two rotating cylinders, i.e.
\begin{equation}
\Omega=a + \frac{b}{R^2}.
\label{Omega}
\end{equation}
With the definition   $\mu = \Om_{\rm out}/\Om_{\rm in}$ and  for $r_{\rm in}=R_{\rm in}/R_{\rm out}=0.5$ one finds $\mu=0.25$.  In order not to conflict with  the Rayleigh criterion for hydrodynamical instability (here $\mu=0.25$), the steep rotation in our calculations is defined by $\mu=0.255$. 
The instability map for the rotational profile with $\mu=0.255$  is given by Fig.~\ref{fig1} for  ${\rm Pm}=1$ and, for reasons of comparison,  for the extremely small value ${\rm Pm}=10^{-6}$. We find that  the dependence of the bifurcation map on the magnetic Prandtl number is very weak. The combination of (differential) rotation and toroidal field is unstable in the open cone between the lower and the upper branches of the non-monotonic function ${\rm Re}={\rm Re}({\rm Ha})$, where the growth rates vanish by definition. The dashed lines for ${\rm Pm}=1$ and ${\rm Pm}=10^{-6}$ mark the locations of the maximal growth rate, which are strikingly close to the lower strong-field line of the instability cone. The eddy viscosities are numerically computed in this instability cone for fixed Reynolds number and various Hartmann numbers, with the general result that $\nu_{\rm T}$ peaks at the location of the dashed line, { i.e.} for maximum growth rates.

\section{Nonlinear simulations}
From Fig. \ref{prandtl} we know that the magnetic Prandtl number in the radiative core of the red giants  varies between 0.1 and 10 so that it makes sense to focus the simulations to $\rm Pm\lsim 1$. 
We first check the validity of the often used relation 
\begin{equation}
\omega_{\rm turn}\simeq \alpha \omega_{\rm gr}, 
\label{turn}
\end{equation}
see the discussion below Eq. (\ref{nuT}). The factor $\alpha$ represents a characteristic Strouhal number formed with the growth rate instead of the characteristic spectrum frequency. The growth rates have been calculated from  linear models  for various Hartmann numbers along the line of maximal growth  in the instability domain as indicated by Fig. \ref{fig1}. The applied rotation law varies between the steep rotation law with  $\mu=0.25$ and the flat rotation law with  $\mu=0.5$ which interval certainly contains the internal stellar rotation.  For the calculation of the turnover rates  we need the nonlinear results for the (vertical) velocity and the cell size $\pi/k$. 

\begin{figure}[t!]
\begin{center}
\includegraphics[width=8cm]{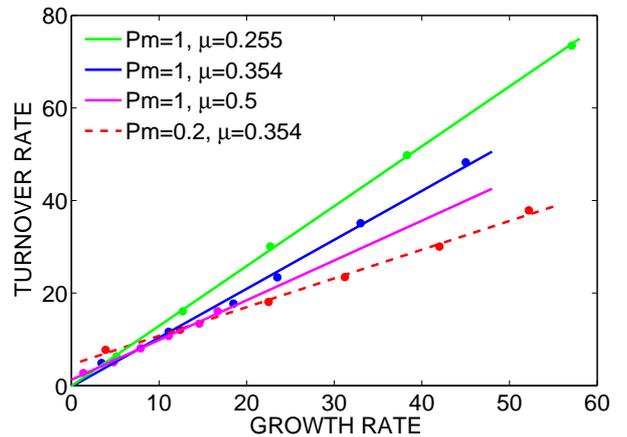}
\end{center}
\caption{The linear relation between the turnover rate and the  growth rate for various $\rm Pm$ and $\mu$. The models possess the maximal growth rates as indicated   in the instability domain of Fig. \ref{fig1}. }
\label{turnover}
\end{figure}

This procedure provides the results plotted in Fig. \ref{turnover} which indeed  provides  the linear relation (\ref{turn}) with $\alpha$ as a  function of  $\rm Pm$ and the shear. As expected, the maximum $\alpha $  (as a modified Strouhal number)  is unity for steep rotation law and large magnetic Prandtl number. It becomes smaller for both decreasing $\rm Pm$ and  for the flatter rotation rates.  
For known $\alpha$ we thus have the puzzling situation that the nonlinear
turbulence intensity $\langle u_z^2\rangle$ can be expressed by the  characteristic quantities $k$ and $\omega_{\rm gr}$ of the linear theory with the simple relation $u_{\rm rms}\simeq \alpha \omega_{\rm gr}/k$.
It does not automatically mean, however, that the effective viscosity can be estimated just  with these quantities as the effect of the magnetic fluctuations cannot be neglected. It might be interesting   to compare the traditional estimate
$\nu_{\rm T}\simeq \alpha^2 \omega_{\rm gr}/k^2$ with the nonlinear results. In any case this relation suggests that the effective viscosity should decline for smaller magnetic Prandtl numbers and flatter rotation profiles. We take for the most optimistic case ($\rm Pm=1$ and $\mu=0.25$) from R\"udiger et al. (2014) that the maximal drift is $0.4 \Omega_{\rm in}$ and the smallest wavenumber is about $3 R_0$ so that the upper limit of the  viscosity due to the hydrodynamic transport can be estimated by $\nu_{\rm T}/\nu\simeq 0.04 \rm Re$ which leads to $\nu_{\rm T}/\nu\simeq 4$ for $\rm Re=100$. We shall see that the nonlinear values do only exceed this value for much higher Reynolds numbers.

\begin{figure}[t!]
\begin{center}
\includegraphics[width=8.5cm]{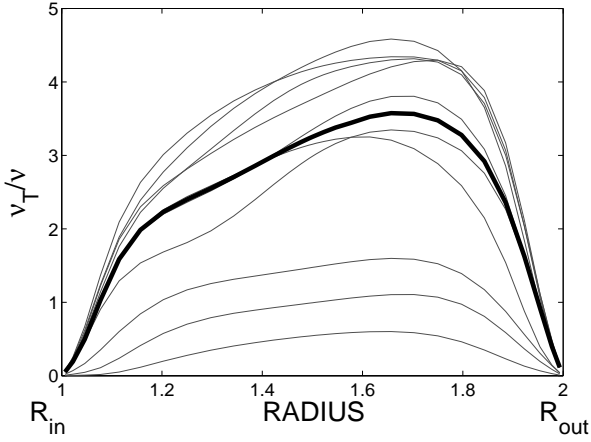}
\end{center}
\caption{Rotation law $\Om\propto 1/R^2$: The fluctuating radial profiles between the inner and the outer radius of the effective viscosity in (\ref{uRuphi}).  The thick solid line gives the temporal average.  $\rm Re=850$.}
\label{fig22}
\end{figure}

Now 
 the effective  viscosity is nonlinearly calculated by computing the RHS of the relation (\ref{uRuphi})
within the instability domain in Fig. \ref{fig1}. For a given Reynolds number, the Hartmann number is varied { until} the maximum value of $\nu_{\rm T}$ is found. Finally, the { maximum viscosity between the inner and the outer cylinder} is taken (see Fig.  \ref{fig22}). The average procedure in (\ref{uRuphi})  concerns { only} the azimuth and the vertical axis. 

\begin{figure}[t!]
\begin{center}
\includegraphics[width=\columnwidth]{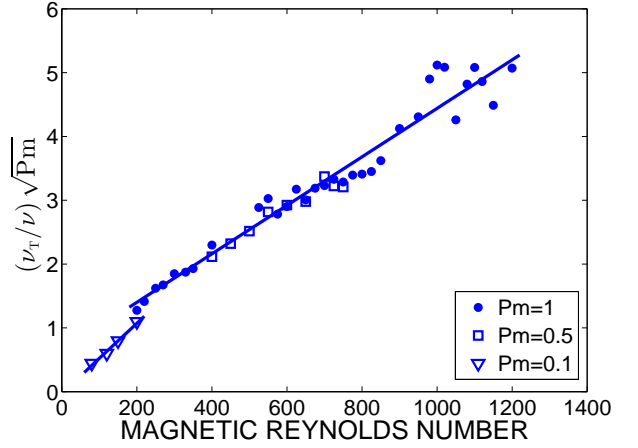}
\includegraphics[width=\columnwidth]{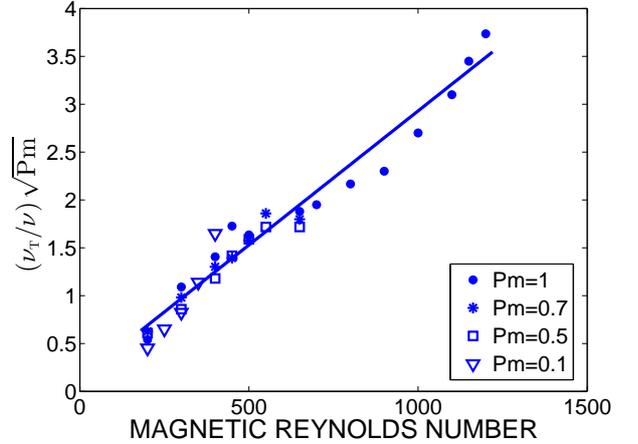}
\end{center}
\caption{Rotation law $\Om\propto 1/R^2$ (top) and $\Om\propto 1/R$ (bottom): The normalized viscosity grows with the magnetic Reynolds number $\rm Rm$ without indication of a saturation ($\rm Pm=0.1...1$, as indicated).
 }
\label{fig31}
\end{figure}

 For various magnetic Reynolds numbers, this procedure yields viscosities which linearly grow for growing ${\rm Rm}$. This is true for all rotation laws between $1/R^2$ and $1/R$ including Kepler rotation (Fig.~\ref{fig31}). For the magnetic Reynolds numbers of the order of 10$^3$ (which our code is able to handle) we do not find any indication of a saturation. One also finds that the resulting viscosity for ${\rm Pm} <1$ scales as $\nu_{\rm T}/\nu\propto  \rm Rm/\sqrt{\rm Pm}$, which can also be written as  
\begin{equation}
{\nu_{\rm T}}\propto \sqrt{\rm Pm}{\Om R^2}
\label{nuT1}
\end{equation}
or, { which} is the same,
\begin{equation}
\frac{\nu_{\rm T}}{\nu}\propto \sqrt{\rm Pm}\ {\rm Re} ,
\label{nuT2}
\end{equation}
with $\rm Re$ the Reynolds number of the fluid. The main result is the rather weak dependence of the viscosity on the magnetic Prandtl number $\rm Pm$.  Obviously -- { since} $\rm Pm$ runs with the electric conductivity -- it is the induction by the radial rotational shear that produces the strong correlations of the magnetic fluctuations given in (\ref{uRuphi}), and this induction vanishes for $\rm Pm\to 0$. 

From Fig. \ref{fig31} it follows that the missing numerical factor  in the relations (\ref{nuT1}) and (\ref{nuT2}) is of order $5\cdot 10^{-3}$. For the solar core, we thus find the rather high value of ${\nu_{\rm T}}/{\nu}\simeq 5\cdot 10^9$. One must take into account, however, that i) our calculations always tend to find the maximal values { of the viscosity} and ii) we { have only verified that} the linear relations (\ref{nuT1}) and (\ref{nuT2}) { hold} for Reynolds numbers up to $10^3$, and cannot account for any saturation effect. We can only assume that the effective viscosity does not exceed the given value.

The dependence of the resulting eddy viscosity on the magnetic Prandtl number has  consequences for stars with  small $\rm Pm$ like  the solar radiative core. There, the eddy viscosity due to the instability of toroidal fields becomes rather small. This is not the case, however, for the radiative cores of red giants with their magnetic Prandtl numbers of order of unity. The magnetic-induced outward angular momentum transport will happen much faster for hot stars in comparison to solar-type stars. 

Inserting characteristic stellar values for $\Om R^2$ one obtains high viscosity values, of the order of $10^{12}$cm$^2$/s. { It is important to note, however,} that in the present paper only {\em  upper limits} for the viscosity are presented, in order to show that the resulting viscosities are not too small. We did not  find the saturation level which is a result of the magnetic feedback onto the rotation laws

\begin{figure}[h!]
\begin{center}
\includegraphics[width=\columnwidth]{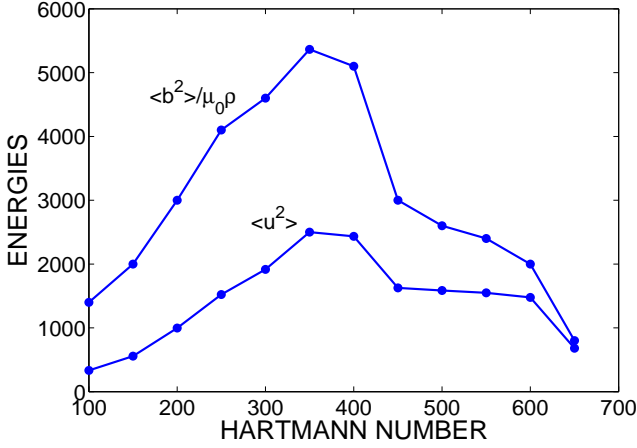}
\end{center}
\caption{Magnetic and kinetic energy for  $\rm Pm=1$ and $\rm Re =800$, normalized { to} the diffusion velocity { squared}.  The energies are in equilibrium { only} for strong magnetic fields ($\mu=0.255$ in this plot).
 }
\label{fig38}
\end{figure}

\begin{figure}[h!]
\begin{center}
\includegraphics[width=\columnwidth]{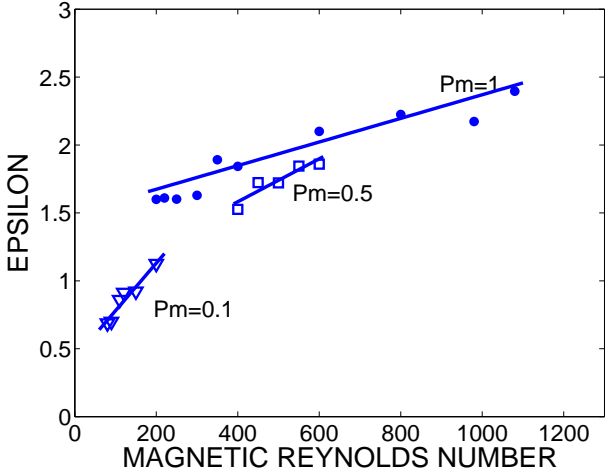}
\includegraphics[width=\columnwidth]{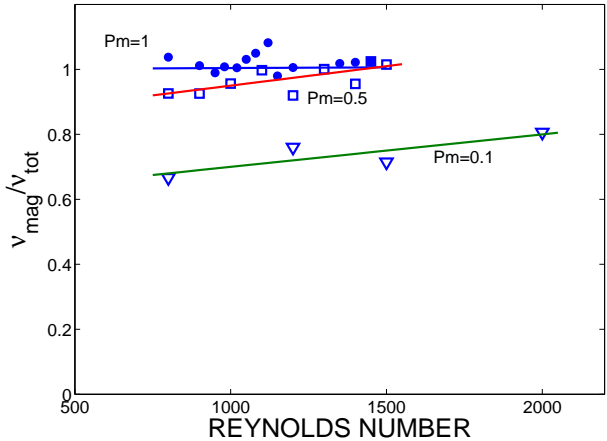}
\end{center}
\caption{Top: Magnetic to kinetic energy $\epsilon$ for various $\rm Pm$, as indicated. The magnetic energy only dominates the kinetic energy for $\rm Pm\gsim 1$. Bottom: The magnetic-induced part in units of the total viscosity ($\mu=0.255$ in this plot).
 }
\label{fig32}
\end{figure}

The role of the Maxwell stress in comparison to the Reynolds stress is { illustrated} in Figs. \ref{fig38} and \ref{fig32}. We find that the magnetic energy dominates the kinetic energy for large magnetic Prandtl numbers, while the energies are almost in equilibrium for small magnetic Prandtl  numbers. 

For a fixed Reynolds number, the ratio $\epsilon$
\begin{equation}
\epsilon=\frac{\langle \vec{b}^2\rangle}{\mu_0\rho\langle {\vec{u}^2}\rangle}
\label{epsil}
\end{equation} 
{ reaches its maximum value} for weak magnetic fields and approaches unity for strong magnetic fields. The kinetic energy never exceeds the magnetic energy (Fig. \ref{fig38}). 
The dominance of the magnetic energy for $\rm Pm\simeq 1$, however, is not too { strong}. It must remain open whether or not $\epsilon$ saturates for very large $\rm Rm$. { For} numerical reasons, the calculations are limited to Reynolds numbers of order $10^3$, where no saturation occurs. Within this interval the ratio $\epsilon$ slightly grows with growing $\rm Rm$.

On the other hand, for large magnetic Prandtl numbers the total viscosity is almost { entirely due to} the Maxwell stress (Fig. \ref{fig32}, bottom).
For large $\rm Rm$ and for $\rm Pm\gsim 0.5$, no visible differences exist { between} the total viscosity and the viscosity due to the Maxwell stress.  The viscosity produced by the instability of current-free magnetic background fields is an almost { purely} magnetic phenomenon. The dominance of the corresponding Maxwell stress over the Reynolds stress { turns out} to be larger than the dominance of the magnetic energy over the kinetic energy. The reason { for this} is that, due to the induction by the radial differential rotation, an almost perfect correlation appears between  the radial and the azimuthal magnetic field fluctuations. { Nevertheless}, as this induction becomes weaker for decreasing magnetic Prandtl number (Fig. \ref{fig32}, bottom)  -- and also for more shallow rotation profiles -- a relation $\nu_{\rm T}\propto {\rm Pm}$  becomes understandable (see Eq. (\ref{nuT2})). For $\rm Pm\to 0$, the resulting eddy viscosity should approach  the small values generated by the Reynolds stress. 

\begin{figure}
\begin{center}
\includegraphics[width=\columnwidth]{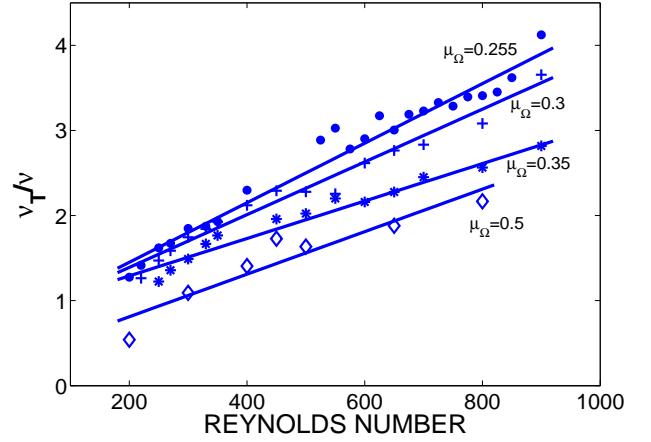}
\end{center}
\caption{The normalized eddy viscosity for $\rm Pm=1$ and the rotation laws $\Om\propto 1/R^2$ ($\mu = 0.25$), $\Om\propto 1/R^{3/2}$ ($\mu = 0.35$), and $\Om\propto 1/R$ ($\mu = 0.5$).
 }
\label{fig4}
\end{figure}

For the special value of $\rm Pm=1$, the influence of the shear on the eddy viscosity is given { in} Fig. \ref{fig4}. There is a {\em nonlinear} influence of the shear on the stress tensor which is often  ignored in the Boussinesq formulation (\ref{uRuphi}). 
As Fig. \ref{fig4} clearly shows, the viscosity is reduced for shallower rotation laws. Obviously, the viscosity itself (and not only the total stress tensor) vanishes for rigid rotation. This { leads to the} important { conclusion that} the diffusive decay of differential rotation is decelerated in time, and the decay time due to magnetic instabilities becomes longer and longer during the diffusion process. The outward transport of angular momentum is thus self-regulated: the original  steep rotation law rapidly decays, producing a flatter rotation law which then decays slower. If the viscosity itself depends on the shear, the considered fluid is called non-Newtonian.  This phenomenon, known to exist for current-driven instabilities (e.g. Spruit  2002), also occurs for the current-free azimuthal magnetorotational instability.

\section{Discussion}
In the sense of a proof of existence, nonaxisymmetric magnetic instabilities under the influence of differential rotation are { shown} to transport angular momentum in a direction orthogonal to the rotation axis. The toroidal background field is chosen as current-free (outside the axis), so that it cannot decay. The mass density is assumed as uniform, suppressing the influence of the buoyancy. The model is thus not applicable to solar type stars { on the main sequence}, but only to the very hot radiative cores of (sub)giants. It is also shown, by use of a numerical stellar model, that the high temperatures in such cores { result in} microscopic magnetic Prandtl numbers varying between $0.1$ and $10$. 
{ This makes it possible to use} a nonlinear MHD code which only works for such Prandtl numbers. The code solves the nonlinear and nonaxisymmetric MHD differential equations in a cylindric setup and provides the $R$-$\phi$-component of the complete stress tensor (Reynolds stress plus Maxwell stress), and averages the resulting number over the azimuth and the axial coordinate.

A spectral element code is used based on the hydrodynamic code of Fournier et al. (2005). 
The solutions are expanded into Fourier modes in the azimuthal direction. The remaining  meridional problems are  solved with a Legendre spectral  method.

If a linear code is used to find the classical eigenvalues for the onset of the marginal instability then all solutions are optimized in the  wave number $k$ yielding  the smallest possible Reynolds 
numbers.

The boundary conditions at the cylinder walls are assumed to be no-slip and perfectly conducting.  The resulting cross correlations thus vanish { at the surfaces of} the two cylinders, { while they have} a maximum { near the} center of the gap between the cylinders (Fig. \ref{fig22}). These { maximum values} are calculated and transformed into viscosity values by means of  Eq. (\ref{uRuphi}). The resulting viscosity values  are compared { for various} Hartmann numbers in the instability cone but for the same Reynolds number (see Fig. \ref{fig1}). If the same procedure is done for various Reynolds numbers and various magnetic Prandtl numbers, one obtains the results presented in Fig. \ref{fig31} for two different rotation laws. Note that the data for different Reynolds numbers do {\em not} belong to the same Hartmann number. The { figures} do not, therefore, contain evolutionary scenarios. As expected, the consequence of the results given in Fig. \ref{fig31} is that the viscosity for given shear  linearly grows with the angular velocity.  The most interesting result, however, is the  dependence of the effective viscosity on the shear: the steeper the rotation law the higher  the viscosity value. Hence, the decay of a nonuniform stellar rotation law is a nonlinear process. The decay time-scale does not remain constant in time as it becomes larger and larger. The non-Newtonian behavior of the magnetized conducting fluid is basically connected with the mechanism of the azimuthal magnetorotational instability which exists only for  differential rotation. It { thus leads} automatically to saturation during stellar spin-down process, which is most violent at its beginning and becomes slower { at later times}. 

As mentioned, the toroidal magnetic field with the current-free radial profile does not dissipate. As we thus fix { the magnetic field} for a given magnetic Prandtl number and consider the dependence of the effective viscosity on the angular velocity, the effective viscosity will grow, passes its maximum close to the line of maximum growth rate shown in Fig. \ref{fig1}, and again sinks to zero at the lowest possible Reynolds number. Figure  \ref{fig31}  demonstrates that, up to Reynolds numbers of the order of $10^3$, along the line of maximum growth rate there is no saturation of the effective viscosities. The stronger the fields, the higher is the { maximum viscosity which can be achieved}. 

While the instability domain is only slightly modified for Prandtl numbers smaller than unity, this is not true for the behavior of the second order correlations. The effective viscosity formed by Reynolds stress and Maxwell stress { decreases} for small $\rm Pm$. It varies by one order of magnitude when $\rm Pm$ varies by two orders of magnitude. The ratio of magnetic energy to kinetic energy (taken for maximum viscosity) { also decreases} for small $\rm Pm$ (Fig. \ref{fig32}, top). Consequently, the Maxwell stress in relation to the Reynolds stress also sinks for small magnetic Prandtl numbers (Fig. \ref{fig32}, bottom). The { effectiveness} of the magnetic perturbations in transporting angular momentum { is thus reduced} for small $\rm Pm$, i.e. with cooler temperatures.

Our results support the conclusion that magnetic instabilities of toroidal magnetic fields in the presence of differential rotation are a viable mechanism to explain angular momentum redistribution in stellar interiors, especially for sub-giant and young red giant stars.

 {}

\end{document}